\documentclass[fleqn,twoside]{article}
\usepackage{espcrc2}
\usepackage{latexsym}
\usepackage{epsfig}
\usepackage{graphicx}
\usepackage[figuresright]{rotating}

\newcommand{\be}{\begin{equation}}
\newcommand{\ee}{\end{equation}}
\newcommand{\ba}{\begin{eqnarray}}
\newcommand{\ea}{\end{eqnarray}}
\newcommand{\<}{\langle}
\renewcommand{\>}{\rangle}

\newcommand{\AmS}{{\protect\the\textfont2
  A\kern-.1667em\lower.5ex\hbox{M}\kern-.125emS}}
\def\vs{{\bf S}}

\def\PRep{{ Phys.\ Rep.\ }}

\title{External field dependence of the correlation lengths
       in the three-dimensional $O(4)$ model
       \thanks{We thank for support by DFG Grant No.\ FOR 339/2-1.}}

\author{J.~Engels, L.~Fromme and M.~Seniuch\\ \vspace{0.5cm}
       {Fakult\"at f\"ur Physik, Universit\"at Bielefeld, \\
        Postfach 100131, 33501 Bielefeld, Germany}}
       
\begin{document}

\begin{abstract}
We investigate numerically the transverse and longitudinal 
correlation lengths of the three-dimensional $O(4)$ model
as a function of the external field $H$. In the low-temperature
phase we verify explicitly the $H^{-1/2}$-dependence of the 
transverse correlation length, which is expected due to the 
Goldstone modes of the model. On the critical line we find
the universal amplitude ratio $\xi^c_T / \xi^c_L = 1.99(1)$.
From our data we derive the universal scaling function for the
transverse correlation length. The $H$-dependencies of the correlation 
lengths in the high temperature phase are discussed and shown 
to be in accord with the scaling functions.\vspace{1pc}
\end{abstract}

\maketitle

\section{INTRODUCTION}
In $O(N)$ spin models with $N>1$ two types of correlation lengths 
appear, corresponding to the transverse and longitudinal spin
components. Like for the magnetization the behaviour of these
correlation lengths in the critical region is described by asymptotic
scaling functions and critical exponents, which characterise the
underlying universality class. In addition, there are predictions
\cite{Fisher:1985,Patashinskii:1973} for the correlation lengths, 
which are related to the presence of massless Goldstone modes. The
measurement of the correlation lengths as functions of the external 
field $H$ enables us to verify these predictions and to determine the
critical parameters and scaling functions. 
We consider the three-dimensional $O(4)$ model because it
is believed to belong to the same universality class as QCD with two
degenerate light-quark flavours at its chiral transition in the
continuum limit. The variant of the model, which we study here, is
defined by
\be
 \beta\,{\cal H}\;=\;-J \,\sum_{<{\bf x},{\bf y}>}\vs_{\bf x}\cdot
\vs_{\bf y} \;-\; {\bf H}\cdot\,\sum_{{\bf x}} \vs_{\bf x} \;, 
\ee
where ${\bf x}$ and ${\bf y}$ are nearest neighbours on a hypercubic
lattice with $L$ points per direction, and
$\vs_{\bf x}$ is a four-component unit vector at site ${\bf x}$.
From the lattice averages $S^{\parallel}$ and $\vs^{\perp}$ 
of the components parallel (longitudinal) and perpendicular (transverse)
to ${\bf H}$ we find
\be
 M \;=\;\<\, S^{\parallel} \,\>~,\quad {\rm and}\quad 
\;\<\, \vs^{\perp} \,\>\;=\;0~,  
\ee 
where $M$ is the magnetization. Correspondingly, there
are two susceptibilities   
\be
\chi_L\;=\; {\partial M \over \partial H}
 \;=\; V(\<\, S^{\parallel2} \,\>-M^2)~,
\ee
\be
\chi_T\;=\; {V \over 3} \<\, \vs^{\perp2} \,\> 
\;=\;{M \over H}~, \quad \mbox{with}\quad V=L^3~.
\ee
The connected correlation functions of the longitudinal and 
transverse spins are defined by 
\be
 G_L({\bf x}) = \<\: S^{\parallel}_{\bf x}
 S^{\parallel}_0 \:\>-M^2~,~
G_T({\bf x})= {1 \over 3} \<\: \vs^{\perp}_{\bf x} \cdot
\vs^{\perp}_0 \:\>\, .
\ee
The large distance behaviour of $G_{L,T}$ is governed by the
exponential correlation lengths $\xi_{L,T}$, except for
$H=0,T\le T_c$. \hfill\break
\indent The spontaneous breaking of the rotational symmetry for
$T<T_c$ gives rise to spin waves or massless Goldstone modes and
leads to the following divergencies for $H\rightarrow 0$ at all fixed
$T<T_c$ 
\be
 \chi_L \sim  H^{-1/2}\:,~ \chi_T \sim H^{-1}\:,~
\xi_T \sim H^{-1/2}\: . 
\ee
The prediction for $\xi_T$ comes from the relation $\xi^2_T
\sim \chi_T$, in accord with the mass interpretation
$m_{\pi}^2=\chi_T^{-1}$ and $m_{\pi}\sim \xi_T^{-1}$; the relation 
$\xi^2_L\sim \chi_L$ (here, $m=m_{\sigma}$) does not hold for $T<T_c$ 
\cite{Fisher:1985}.

\section{NUMERICAL RESULTS}

Our simulations were done on lattices with periodic boundary conditions
and linear extensions $L=48,72,96$ and 120. We used the Wolff single 
cluster algorithm and made 20000 measurements for each fixed $(H,J)$-pair.
Between two measurements 100-3000 cluster updates were performed. In order
to determine the correlation lengths we first calculated spin averages
over planes and their respective correlation functions $\bar G(\tau)$.
From the correlators at $\tau$ and $\tau +1$ we then derived an 
effective correlation length $\xi_{eff}(\tau)$. When $\xi_{eff}(\tau)$
reached a plateau inside its error bars we used the corresponding
$\tau$-range to find $\xi$ from a fit to
\be
 \bar G(\tau) =A[\exp(-\tau/\xi)+ \exp(-(L-\tau)/\xi)] ~.
\ee

\subsection{Critical behaviour}

 In the thermodynamic limit ($V\rightarrow \infty$)
      and close to $T_c$ critical observables show power law behaviour in
      the reduced temperature $t=(T-T_c)/T_c$ : For $H=0$
      \ba
      M\!\!&=&\!\! B (-t)^{\beta}~,~\! \quad t<0\\ 
      \chi_L\!\!&=&\!\! C^+ t^{-\gamma}~,~\, \quad t>0\\ 
      \xi\!\!&=&\!\! \xi^+ t^{-\nu}~,~~ \:\quad t>0 
      \ea
      and at $T=T_c$ for $H>0$
      \ba
 M\!\!&=&\!\! B^c H^{1/\delta}~, \label{mpow}\\ 
 \xi_{L,T}\!\!&=&\!\! \xi^c_{L,T} H^{-\nu_c}~,\quad \nu_c=\nu/\beta\delta~.
 \label{xipow}
      \ea
Here, the temperature $T$ is the inverse of the coupling 
$J=1/T$ and $J_c=0.93590$. From \cite{Engels:1999wf} we use the values
$B=0.9916(5)$ and $\beta=0.380$. We have fitted the data for $M$ 
to Eq.\ (\ref{mpow}) and find $ B^c\;=\;0.721(2)$ and
$\delta \;=\; 4.824(9)$. From $\beta$, $\delta$ and the hyperscaling 
relations all other exponents are fixed. In Fig.\ 1 we show the $H$-dependence
of $\xi_L$ and $\xi_T$ at $T_c$ and compare it to a fit to Eq.\ 
(\ref{xipow}), with the critical amplitudes    
\be
 \xi_T^c\,=\,0.838(1)~,~\xi_L^c\,=\,0.421(2)~,
\label{critxitc}
\ee
which leads to the universal ratio $U_{\xi}=\xi_T^c/\xi_L^c=1.99(1)$,
a result expected actually below $T_c$ from a relation between $G_L$
and $G_T$ \cite{Fisher:1985,Patashinskii:1973}. At $H=0$, $T>T_c$, where
$\xi_T=\xi_L=\xi$ and $\chi_T=\chi_L$ we have made similar fits to our
data and find the amplitudes $\xi^+=0.466(2)$, $C^+= 0.231(2)$. From the
\begin{figure}[t]
\begin{center}
   \epsfig{bbllx=63,bblly=265,bburx=516,bbury=588,
        file=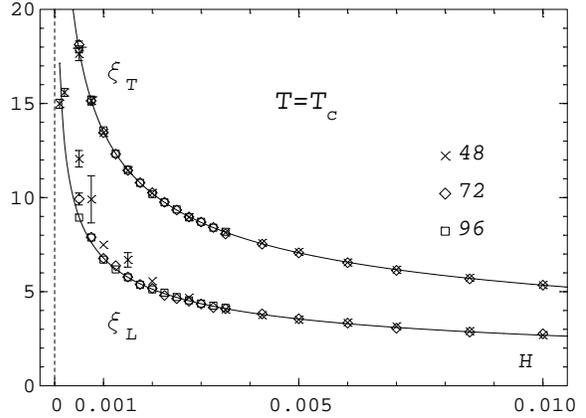,width=70mm}
\end{center}
\vspace{-0.9cm}
\caption{The correlation lengths $\xi_T$ and $\xi_L$ at $T_c$ versus
$H$ from $L=48,72$ and 96-lattices. The lines are fits to Eq.\ (\ref{xipow})
with the result (\ref{critxitc}).}
\label{fig1}
\vspace*{-0.6cm}
\end{figure}
measured amplitudes we obtain more universal amplitude ratios 
\cite{Pelissetto:2000ek}
    \ba
R_{\chi} \, =\!\!&C^+ (B^c)^{-\delta} B^{\delta-1}\,=\, 1.084(18)~,\\
Q_c\,=\!\! &B^2 (\xi^+)^d /C^+\,=\,0.431(9) ~. 
    \ea
\vspace*{-0.8cm}
\begin{figure}[hb!]
\begin{center}
   \epsfig{bbllx=51,bblly=265,bburx=516,bbury=588,
        file=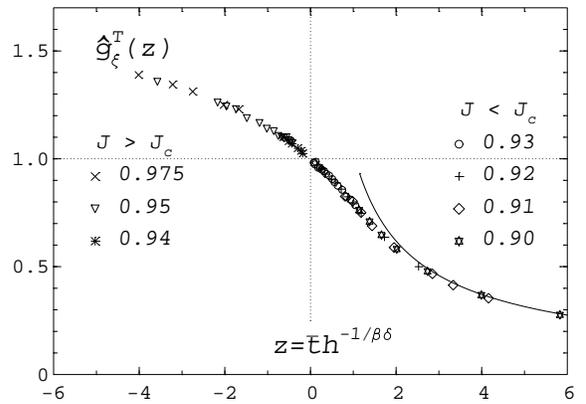,width=70mm}
\end{center}
\vspace{-0.9cm}
\caption{The normalized transverse scaling function 
${\hat g}_{\xi}^T(z)= \xi_T h^{\nu_c}/g_{\xi}^T(0)$ for various 
$J$-values. The line is the asymptotic form, Eq.\ (\ref{asyt}).}
\label{fig2}
\vspace*{-0.7cm}
\end{figure}
\subsection{The scaling functions}

In the critical region the dependence of the observables on the
temperature and the external field is described by scaling functions
    \ba
     M\,=\!\!&h^{1/\delta} f_G (z)~,\!\!\! &\chi_L\,=\,
     h^{1/\delta -1} f_{\chi} (z)/H_0 ~, \\
     \xi_T\,=\!\!&h^{-\nu_c}\,g_{\xi}^T (z)~,\!\!\! &\xi_L\,=\,h^{-\nu_c} 
     g_{\xi}^L (z)~, 
     \ea
where $z\,=\, {\bar t} h^{-1/\beta\delta}$, and ${\bar t}=t B^{1/\beta}$,
$h=H/H_0$ with $H_0=(B^c)^{-\delta}$, are the normalized reduced temperature
and field. The asymptotic behaviour of $g_{\xi}^{L,T} (z)$ for
$z\rightarrow\infty$ ($H\rightarrow 0,\: t>0$), is
\be
 g_{\xi}^{L,T} (z)  \; {\raisebox{-1ex}{$\stackrel 
   {\displaystyle =}{\scriptstyle z \rightarrow \infty}$}} \;
   \xi^+ B^{\nu/\beta} z^{-\nu}~. \label{asyt} 
\ee
In Fig.\ \ref{fig2} we show the normalized (and universal)
scaling function of the transverse correlation length. Whereas it was 
always possible to determine $\xi_T$, we could only determine $\xi_L$
above $T_c$, because below $T_c$ a whole spectrum of states contributes
to the longitudinal correlation functions. 
As it turned out, the shape of ${\hat g}_{\xi}^T(z)$ is similar to
$f_G(z)$ and - at least for $z>0$ ($T>T_c$) - the shape of 
${\hat g}_{\xi}^L(z)$ resembles that of $f_{\chi}(z)$ (for example,
the peak positions are the same). This similarity comes from the 
relations $\xi_T^2 \sim \chi_T$ and $\xi_L^2 \sim \chi_L$. For more details 
see Ref.\ \cite{Engels:2003nq}.

\subsection{The $H$-dependence at fixed $T<T_c$}
 
\begin{figure}[b!]
\begin{center}
   \epsfig{bbllx=63,bblly=280,bburx=516,bbury=538,
        file=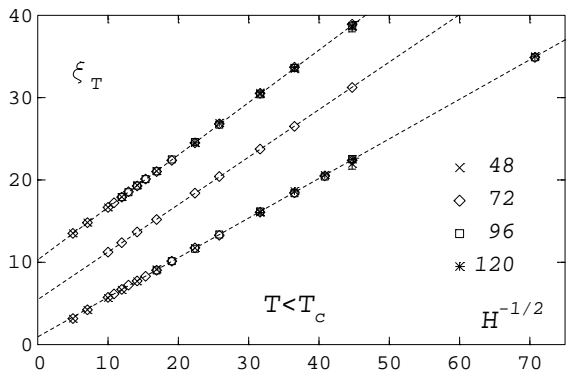,width=70mm}
\end{center}
\vspace{-0.9cm}
\caption{$\xi_T$ in the low temperature phase at $J=1.0,\,0.975$ and 0.95
vs. $H^{-1/2}$. The results for $J=1.0(0.975)$ are shifted upwards by
10(5).}
\label{fig3}
\vspace*{-0.5cm}
\end{figure}

In Fig.\ \ref{fig3} we show our results for $\xi_T$ at fixed $T=1/J$ below
$T_c$. The lines are fits to the form expected due to the presence of
{\em Goldstone modes}
\be
\xi_T\,=\,x_0 H^{-1/2} +x_1 ~.
\label{puregold}
\ee
We see a clear verification of this prediction. The increase
of the slope $x_0=0.481(2),\,0.577(1)$ and 0.637(2) for the three 
$J$-values with decreasing temperature is in accord with
     the scaling function. By comparing the asymptotic behaviour of
     $g_{\xi}^T$ to the $H$-dependence of $\xi_T$ we find
\be 
x_0 \sim (-\bar t)^q~,~ \mbox{with}
     ~q\,=\,-\nu+\beta\delta/2\,=\,0.1789~.
\ee
Indeed, the ratios between the different $x_0$ confirm this formula.
\begin{figure}[t]
\begin{center}
   \epsfig{bbllx=63,bblly=265,bburx=516,bbury=588,
        file=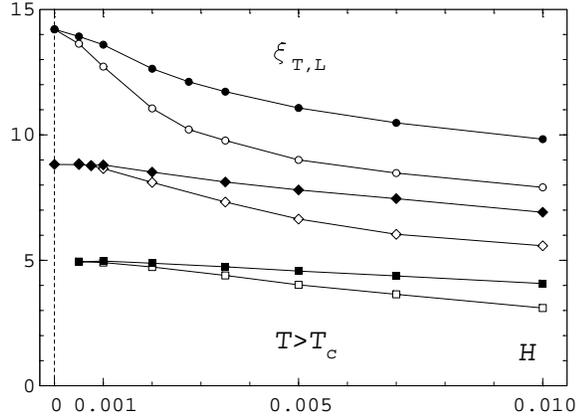,width=70mm}
\end{center}
\vspace{-0.9cm}
\caption{$\xi_T$ (filled) and $\xi_L$ (empty symbols) for
 $J=0.92$ ($\circ$), 0.91 ($\Diamond$) and 0.90 ($\Box$). The results
 for $J=0.92\,(0.91)$ are shifted upwards by 5\,(2.5).}
\label{fig4}
\vspace*{-0.5cm}
\end{figure}

\subsection{The $H$-dependence at fixed $T>T_c$}
 
 Above $T_c$, we have for all $H>0$ the inequality $\xi_T>\xi_L$,
 whereas at $H=0$ the correlation lengths coincide. In Fig.\ \ref{fig4}
 we show $\xi_T$ and $\xi_L$ at $T>T_c$ versus $H$. With increasing 
 temperature the curves become flatter. The behaviour can again be
 explained from the asymptotic scaling function. Since $\xi$ is an
 even function of $h$ we must have
\be
 \xi \,=\, {\bar t}^{-\nu} [ g_0 + g_1 {\bar t}^{-2\beta\delta}h^2 
 + \dots]~, 
\ee 
 with constant $g_i$. Yet, close to $T_c$, where $|z|$ is small, but not
 $h$, the expansion of $g_{\xi}$ in powers of $z$ leads to a different 
 $h$-dependence
\be
 \xi\,=\, h^{-\nu_c}[g_0^c + g_1^c {\bar t} h^{-1/\beta\delta}
  +\dots]~.
\ee

\end{document}